# A systematic review of guidelines for the use of race, ethnicity, and ancestry reveals widespread consensus but also points of ongoing disagreement

Madelyn Mauro[1], Danielle S. Allen[1], Bege Dauda[2,3], Santiago J. Molina[4], Benjamin M. Neale[5,6,7], Anna C. F. Lewis[1,8*]

1 Edmond J Safra Center for Ethics, Harvard University, Cambridge, MA, USA.
2 Center for Global Genomics and Health Equity, University of Pennsylvania, Philadelphia, PA, USA.
3 Institute of Clinical Bioethics, Saint Joseph's University, Philadelphia, PA, USA.
4 Department of Sociology, Northwestern University, Evanston, IL, USA
5 Broad Institute of Harvard and MIT, Cambridge, MA, USA.
6 Analytic and Translational Genetics Unit, Massachusetts General Hospital, Boston, MA, USA
7 Center for Genomic Medicine, Massachusetts General Hospital, Boston, MA, USA.
8 Division of Genetics, Department of Medicine, Brigham and Women's Hospital, Boston, MA, USA.
* Corresponding author

**ABSTRACT**

The use of population descriptors like race, ethnicity, and ancestry in science, medicine and public health has a long, complicated, and at times dark history, particularly for genetics, given the field's perceived importance for understanding between-group differences. The historical and potential harms that come with irresponsible use of these categories suggests a clear need for definitive guidance about when and how they can be used appropriately. However, while many prior authors have provided such guidance, no established consensus exists, and the extant literature has not been examined for implied consensus and sources of disagreement. Here we present the results of a systematic review of published normative recommendations regarding the use of population categories, particularly in genetics research. Following PRISMA guidelines, we extracted recommendations from n=121 articles matching inclusion criteria. Articles were published consistently throughout the time period examined and in a broad range of journals, demonstrating an ongoing and interdisciplinary perceived need for guidance. Examined recommendations fall under one of eight themes identified during analysis. Seven are characterized by broad agreement across articles; one, *Appropriate definitions of population categories and contexts for use*, revealed substantial fundamental disagreement among articles. While many articles focus on the inappropriate use of race, none fundamentally problematize ancestry. This work can be a resource to researchers looking for normative guidance on the use of population descriptors, and can orient authors of future guidelines to this complex field, contributing to the development of more effective future guidelines for genetics research.

**INTRODUCTION**

Evidence of race-based health disparities has mounted in recent years, especially during the COVID-19 pandemic. [1-5] As researchers have stressed the role of structural racism in generating these disparities [6] and the importance of racially-stratified data to developing a better understanding of them[7], prominent institutions like the National Academies of Sciences, Engineering, and Medicine (NASEM) and the National Human Genome Research Institute have begun to wrestle with questions of the usefulness of population descriptors. [8,9] A recent series of commentaries in the New England Journal of Medicine illustrates the significant disagreement on the value of race as a variable in biomedicine, as well as a turn towards concepts from genetics as a potentially suitable alternative. Vyas et al. argued that the insertion of race into clinical tools relies on faulty assumptions about the genetic contribution to racial categories and can lead to interpretations of racial disparities as unavoidably genetically determined when they may be avoidably socially determined.[10] Borrell at al. raised the counterclaim that leaving awareness of race and ethnicity out of healthcare can exacerbate racial and ethnic disparities by failing to monitor or condemn them, and argued that race and ethnicity should therefore be used alongside genetic ancestry to understand health outcomes. [11] Oni-Orisan et al. highlighted a potential contribution of genetic ancestry to COVID-19 disease outcome, and after broadly reviewing the contribution of genetic difference to health disparities and the overlap between genetic and racial diversity, proposed that "the ultimate goal… would be to replace race with genetic ancestry in an evidence-based manner." [12] But this turn to genetic ancestry as a more objective way to capture biological difference between groups has its own pitfalls.[13,14] Complex genetic ancestry information is most often smoothed into continental ancestry categories: a dangerous oversimplification due to the striking resemblance of continental ancestry categories to racial categories.

Given the complicated nature of this topic and this debate, a researcher considering employing population categories in their work might search for guidelines to consult in order to determine which population categories to use for which purposes. They would find a plethora of articles offering such guidelines, but with a wide range of focuses and varying levels of specificity; this phenomenon suggests a pervasive need for guidance in the scientific and medical community, but also highlights a continued lack of explicit, centralized normative guidance..

Prior works that have assessed this complicated body of normative literature have either examined only small sets of recommendations or have analyzed the impact of a single recommendation or set of recommendations on the practices of authors and/or clinicians. [15–17] A systematic review of all existing normative guidelines is lacking. In its absence, it is impossible to identify the recommendations that have been echoed by multiple authors, representing areas of normative consensus, or to pinpoint areas of consistent disagreement. The former is useful because commonly-provided recommendations point to noncontroversial areas of improvement for researchers, clinicians, and public health practitioners, and can also influence future guidelines by highlighting the topics and sentiments that have already been well-expressed in the field. The latter point to topics that merit further examination and likely hold some of the most pressing ethical concerns that underlie the employment of population categories in genetics and across scientific fields.

Here we present the results of a systematic review of normative guidelines relating to the use of population categories in science, medicine, and public health, thus providing an overdue clarification and categorization of normative works in this field. It focuses particularly on the relevance of genetics to the use of these categories, given the perceived importance of genetics for understanding between-group differences. [10-12] This systematic review can provide a resource for consultation by researchers, funders, practitioners, regulatory bodies, and others who may find themselves lost in this otherwise disordered and overwhelming space. It also gives the authors of future guidelines — such as the recently convened National Academic committee on this topic — a much-needed orientation to existing work. This piece thus contributes to informing and improving future work and to pushing the field towards clearer and more effective future sets of guidelines.

**METHODS**

*Search Method*

This is a systematic review of literature containing normative recommendations for the use of race, ethnicity, and ancestry in science, medicine and public health, with particular focus on genetics. We conducted the review in accordance with guidelines established by Preferred Reporting Items for Systematic Reviews and Meta-Analyses (PRISMA).[18] In December of 2021, two different search methods were applied in parallel to identify and collect articles of interest. The PubMED electronic database was searched using the following string: "population groups AND genetics AND human research AND bioethics[sb]." The [sb] qualifier limits the set of articles returned by the search to only include those labeled by PubMed as pertaining to bioethics. "Population groups" was used rather than "race AND ethnicity AND ancestry" because the phrase is a MeSH term—a database-specific phrase that more thoroughly and specifically refines searched articles. At the same time, a search was conducted using the Google search engine with the string "race, ethnicity, and ancestry in genetics". The first 100 articles of the results of this search were included. The Google search was designed and included to ensure that articles accessed by authors conducting a preliminary search for normative recommendations to guide their work would be included. Given the transformative impact of the Human Genome Project, which was completed in the first few years of the 21st century, both searches were refined to only include articles published since January 1st, 2000.

This combination of search terms yielded ten of thirteen articles previously identified by one reviewer as relevant to the topic of this systematic review. Although other tested combinations yielded more of these articles, they raised the total number of returned articles to the tens or hundreds of thousands—thus, in an effort to balance the aim of including these articles with the aim of maintaining precise focus within this review, the above-mentioned search term combination was chosen.

*Preliminary screening criteria*

In order to refine the results of this search to yield articles with a normative focus, the titles and full abstracts of all returned articles were read. Each article was screened for how likely it seemed to contain normative recommendations on the use of race, ethnicity, and ancestry in science, medicine, or public health. Gray literature, such as transcribed presentations and speeches, as well as pieces published in popular media were not excluded. Standards for inclusion were established by comparing the labeling of the first 100 PubMED articles between two reviewers .The rest of the articles were labeled by one reviewer. Figure 1 illustrates the full process of inclusion and exclusion.

*Extraction*

Normative recommendations from these articles were extracted using a three stage process. First, summarized normative recommendations were extracted from the abstract and conclusion. Second, if articles contained tables or sections explicitly allocated to reporting recommendations, the contents of those tables and sections were extracted. Finally, each article was searched for the following words and phrases to identify further normative recommendations for extraction: *recommendations*, *guidelines*, *should*, *must*, *need*, *ought*, and *consider*. This strategy was determined to be sufficient by comparing, for 10 articles, recommendations identified using this strategy to recommendations extracted by another reviewer based on an examination of the whole text. Along with text of each recommendation, any accompanying text pertaining to the justification of the recommendation was extracted.

Additionally, the following basic data about each article were collected: authors, all countries of author affiliation and country of majority author affiliation, journal, country of journal publication, publication year, and number of citations as recorded by PubMed. For the purpose of later categorization, journals were sorted into six categories—science, medicine, public health, law, ethics, or other. Finally, articles were tagged if their explicit aim was to provide normative recommendations, and if they pertained to ancestry.

After all the recommendations were extracted, thematic content analysis was conducted by examining each recommendation and identifying emerging themes. In order to establish the list of themes, two reviewers examined the recommendations from several articles. Theme assignments were all made by one reviewer. Initial themes were grouped into broader themes. Another reviewer assessed theme assignments and any disagreements were discussed and resolved.

**RESULTS**

*Study Selection and Characteristics*

As shown in Figure 1, this search yielded 1,073 articles in total after duplicates were removed, 1,035 of which emerged from the PubMED search and 38 of which emerged uniquely from the Google search. 218 articles remained after preliminary screening, and 121 were included in the final analysis. 384 normative recommendations were extracted from these 121 articles.

There is a clear dominance of articles published in the first decade of the 21st century, surrounding the completion and publication of the Human Genome Project, as shown in Figure 2a—however, Figure 2a also shows that there have been sustained publications on this topic throughout the examined time period. The articles examined are broadly distributed across journals with different focuses, although most articles hail from scientific or medical journals, see Figure 2b.

Most if not all contributing authors for 102 articles (84% of total article yield) were affiliated with the USA. Four articles hailed from Canadian authors, three from authors in Germany, two from authors in the U.K., two from Italian authors, and one each from authors in Austria, Iceland, Japan, Scotland, Singapore, and South Africa. The country of majority authorship remains undefined for two of the included articles, one of which is an editorial with no explicit author, and one of which was written by four authors each from different countries.

Although many recommendations called for the formation of working groups to publish research or normative commentary on the use of race, ethnicity, and ancestry in the sciences, only three of the included articles themselves were written on behalf of a group. The articles have a median of 27.5 citations, with an interquartile range of 10.75 to 65.25, and with 23 of the articles having over 100 citations and two having over 500. Finally, 34 of the 121 examined articles contain a total of 60 recommendations that mention ancestry.

*Thematic Discussion*

Twenty-eight distinct themes (referred to as sub-themes from here on) were identified and grouped further into eight broader themes. Table 1 lists these themes and their associated sub-themes, as well as the articles in which recommendations pertaining to these themes and sub-themes were found. Recommendations that did not fall under any of these themes were marked "Other". Not all sub-themes will be fully addressed in the discussion of each theme: refer to Table 1 and to the data listed in Supplementary Materials for the complete list of recommendations and the sub-themes assigned to them.

*1. The need for transparency*

Forty-seven (12.2%) recommendations extracted from the analyzed articles propose a need for transparency from researchers conducting studies that employ population categories. Almost half of those urge investigators and authors to be transparent about *how* any population categories used in their studies are defined. This includes how participants were categorized, and if they were categorized by the investigators or by the participants themselves. [19–27] Many recommendations also encourage authors to provide their own definitions of the broad population categories they use (e.g., of the term "race"), often citing the confusion that results from leaving population categories undefined and the fact that many population categories are assumed to be potentially explanatory variables without justification.[15,24,28–31] Five of the recommendations in this sub-category pertain specifically to ancestry. The conflation of population categories with biological similarity is addressed by two of them, which urge researchers to acknowledge when they are using categories as a proxy for ancestry and to note explicitly how much biology is assumed to contribute to the categories they employ. [24,32] The other three all belong to one article, which urges investigators to clearly differentiate race, ethnicity, and ancestry when using more than one of these in a study and to fully describe the methods by which genetic ancestry was determined or inferred. It also calls for investigators not just to use the term ancestry, but to specify "genetic ancestry" or "inferred genetic ancestry" based on the methods used to assign ancestry labels. [15]

The second most-popular sub-category of recommendations within this theme is made up of recommendations that urge investigators and authors to be transparent about *why* they have employed population categories in their research. These mainly encourage investigators to make explicit the relevance of any employed population categories to their research question(s), especially when the category employed is race. [22–24,26,32,33] Five of the recommendations within this subcategory more specifically urge authors to justify the use of non-genetic, non-biological population categories either as additional variables in genetic studies or in place of genetically-inferred variables. [22,24,30,32]

The rest of the recommendations that fall under this theme relate to transparency about data, collection methods, and the link between results and conclusions. Six recommendations call for investigators to make their collection methods and all of their data publicly accessible, so that other researchers can further investigate the validity of any claims made on the basis of this data and so that investigated communities can access study results.[33–38] Six more urge investigators to fully analyze and explain their findings: all of these recommendations share the goal of discouraging unsubstantiated inferences from data and treating every finding carefully, either by proposing follow-up studies to investigate the finding further, exploring the potential contribution of non-population factors, or simply making sure not to report observed associations without any comment on potential nuance.[16,23,39]

*2. Awareness of impact on particular communities*

Seventy-five (29.4%) of all extracted recommendations fell under this theme. Most of these encourage awareness within the scientific and medical communities of the past and present impacts of racism in science and medicine. Besides compelling researchers to be generally wary of group harms, misconceptions, and mistrust that can be seeded by irresponsible race-based research [40], many of these recommendations advise researchers not to repeat the grave mistakes made and harms done throughout the racist history of science and medicine. [25,28,32,41,42,42–46] All of these articles encourage developing awareness of these historical lessons through the education of investigators and authors. However, while some identify past mistakes and the lessons that should be learned from them—these include Nazism and the Tuskegee Syphilis Study to caution against "eugenic temptation" [42] and the use of racial labels to insinuate inferiority or superiority [32]—many simply broadly advise that research today should be historically informed. [28,43,45] A separate class of recommendations within this sub-theme encourages education of clinicians and researchers about the ways in which racism still extends its hand into science and medicine today. [25,32,41,44,46] Finally, several recommendations urge clinicians to acknowledge the impact of racism on their practices, so that they can root it out and develop more sophisticated understandings of race. [23,47–50]

Thirty-three recommendations within this theme encourage investigators to respect the study populations examined in their work. Most of these address the concern that population-specific studies are often conducted without proof that a population-specific focus has tangible scientific or medical benefit, and so urge either review boards, the FDA, or society as a whole to ensure that this is true for all research involving only particular populations. [11,51–53] Others advocate for improving the sensitivity of the informed consent process for minority populations by implementing measures to ensure all participants have a holistic understanding of the research question and scope, and of all of the potential uses of any collected material.[51,54–56] Thirteen more encourage researchers to foster a respectful partnership with studied communities by releasing their eventual conclusions to the studied communities [23,38,52,54,56,57] and maintaining respectful communication with them throughout and beyond the study [37,53,56,58–60]. Relatedly, twelve recommendations within this theme urge investigators to consult their studied populations throughout the research design and implementation process in order to establish a respectful partnership, develop trust, ensure those populations are appropriately named in the study, and identify risks that may not be obvious to the researchers themselves or to review boards. [41,53,55,56,58,61–63]

*3. The use of appropriate statistical methodologies*

Forty (10.4%) of the extracted recommendations fell under this theme, and many of these urge investigators to be aware of and account for the contribution of factors other than race or ethnicity to health outcomes when designing and conducting research. These factors include racism [64], socioeconomic status [21,26,51,64–67], environmental exposures [51,64,68], education [21,26,51,64], place of residence [26,64], ancestry [51,65,67,69] and cultural practices [65,66,68,69]. One of these articles argues that an examination of the effects of ancestry (as opposed to race or ethnicity) on health should also include a consideration of other factors [68]. Recommendations from one article emphasized that the hypothesis being tested should dictate which of race, ethnicity, or ancestry should be used [23].

Another subset of recommendations within this theme discourages investigators from making causal claims based on associations found in their data. These recommendations share the central message that correlation does not and should not imply causation, and that researchers should interpret their data carefully before using them to make and support claims. [11,24,25,36,43,66] Three of these have a more specific focus: one denounces the use of genetic data to enforce between-group differences in any context [25], and two others caution against overemphasizing the impact of ancestry on health [66,70].

Finally, some recommendations pertaining to statistical methodologies set out specific guidelines for the statistical interpretation of genetic data. These include establishing sufficient statistical significance in any investigated associations between racial, ethnic, genetic, or other categorization and phenotype [24], conducting pooled rather than stratified analysis to avoid "arbitrary clustering decisions" [32], how to select the number of principal components needed in principal components analysis (PCA) to account for population stratification [32], and incorporating socio-historical considerations into understanding population-specific genetic variation [71].

*4. Public reactions and public engagement*

Most of the thirty (8.9%) recommendations within this theme encourage anyone investigating links between population categories and health to be aware of all possible interpretations of their conclusions, and to present findings in a way that minimizes dangerous misinterpretations. This category shares a concern about racism with the recommendations in the category *Awareness of impact on particular communities*, but this category places emphasis on public misconceptions rather than specific harms done to communities. Those aimed at the practices of investigators urge them to note the limitations of their methods [62], avoid overstatement or generalization in their presentation of results [72,73], make accurate versions of their conclusions accessible to popular media to avoid sensationalization [51,74,75], and correct any observed misinterpretations of their results [36]. Five recommendations specifically state that researchers and reviewers should anticipate any potential socioethical problems raised by the associations such research makes or could make, such as the reinforcing or creation of social stigma, and address them within the research or use them as a basis for rejection of funding or publication. [15,36,51,76,77]

    The recommendations aimed at healthcare advertising and direct-to-consumer ancestry testing companies also encourage those formulating public messaging to be aware of and attempt to prevent potential misinterpretations of their claims [78–80]. They recommend that direct-to-consumer ancestry companies demonstrate how consumers should interpret their results and explain to consumers the difference between ancestry and race, ethnicity, and other group membership. [80] Messaging from healthcare advertising, meanwhile, should be careful not to suggest links between race, ethnicity, ancestry, and health that could legitimize racism. [79] One recommendation explicitly states that this messaging should avoid insinuating any biological basis for racial categories.[78]

    The rest of the recommendations within this theme encourage investigators and institutions to increase the involvement of the public in the refinement of future population-based research. The two main ways in which they recommend this be accomplished are through consultation of the public throughout the research process [23,40,81] and through education efforts that raise awareness of racism and its impacts and of the pitfalls of current research involving population categories and health. [36,82–84]

*5. The need for diverse samples and practitioners*

    This theme consists of nineteen (5%) extracted recommendations that bring attention to the need for diversity both in research studies and in the medical and scientific community as a

whole. Recommendations that encourage more diverse recruitment of study participants mostly advocate for increased recruitment of minority populations, either to better represent global diversity [21,67,85–87] or, in the context of ancestry, to improve the applicability of clinical genomics as a whole to those of non-European genetic ancestry [68,88,89]. Three recommendations lie tangential to this sub-theme, as they share the motivation of refining the true applicability of population-based research: they urge researchers to note explicitly how representative their study participant cohorts are of the larger population(s) being studied. [22,24,86]

The last seven recommendations within this theme call for the makeup of and discussions within the scientific and medical community to be diversified: according to these recommendations, doing so further incorporates those with an interest in improving population-focused research into medicine and science and develops a wider understanding of and appreciation for diversity. [38,50,59,74,86,90]

*6. The need for an appreciation of nuance*

Twenty-seven (7%) extracted recommendations fell under this theme, which consists of recommendations that encourage researchers and/or clinicians to consider and use race, ethnicity, ancestry, and/or population in a more nuanced fashion.

Twelve of these specifically pertain to race and ethnicity. Five encourage researchers and clinicians to acknowledge that race and ethnicity are not strict categories and are influenced by multiple circumstances. [23,28,48,81,91,92] Two others offer specific variables that may provide useful replacements for the category of race specifically—ethnicity [21] and geographic origin [93]. The last three simply urge scientists and clinicians to acknowledge the bounds of the relevance of racial and ethnic categories. [27,94] Four of these touch on conceptions of ancestry and population; they broadly propose that population categorization should not be immediately assumed to be legitimate, and that its justification, the types of categories that currently exist, and the ways in which people may be assigned to categories should consistently be questioned. [43,93,95]

Finally, nine of these encourage those in the scientific and medical community to consider the relationship between race, ethnicity, ancestry, and population with more nuance. Seven recommendations in this category encourage that any associations between biology and existing categories be made only upon findings of substantial correlation, clear justification, and unbiased presentation. [24,52,90,96–98] Another calls for a reexamination of the historical contributions to genetic variation, so that we may understand why it is sometimes customary to assign biological significance to arbitrary categories. [95] The final recommendation within this theme calls for researchers to use a specific algorithm developed by the authors that attempts to refine self-identified race and ethnicity. [99]

*7. Appropriate definitions of population categories and contexts for use*

A subset of the eighty-three (24.9%) recommendations within this theme set out how certain population categories should or should not be defined. Unlike the other identified themes, this theme comprises many mutually incompatible recommendations. Many of these encourage researchers to define race not as a biological phenotype, but as a complicated social phenomenon, or to categorize populations by socio-environmental variables instead of by race. [23,48,78,81,100,101] However, one of these introduces the social concept of race not as a replacement for a biological concept of race, but as a definition of race that should be considered in concert with a biological definition of race [100]. Others are concerned with how strictly population categories should be defined: while one states that populations should only be defined by genetic variation [93], another considers rare genetic variants to be too small and strict of a basis for population categorization [43]. Another suggests that populations should be defined in multiple ways [95], while still another maintains that population categories should not and cannot be broadly defined, and urges researchers to instead circumstantially define them [28]. Finally, some provide their own definitions of race, ethnicity, and/or ancestry rather than commenting on how they should be defined [52,67]. For example, in 2015 Mersha and Abebe proposed, *"To better understand human genetic variation in the context of health disparities, we suggest using "ancestry" (or biogeographical ancestry) to describe actual genetic variation, "race" to describe health disparity in societies characterized by racial categories, and "ethnicity" to describe traditions, lifestyle, diet, and values."*

A second subset unconditionally approves of the scientific and medical communities investigating links between population categories and health. Some claim that self-identified racial categories overlap with genetic groupings and so provide a more feasible way to establish ancestry than empirical genotyping [102]; others justify this position by claiming that race- and ethnicity-based disparities still exist, and that it is fruitless to try to eliminate them and understand their causes without examining them [23,28,96,103–105]. Three more recommendations focus on the value of using race or ethnicity to examine the effects of racism on health [95,104,106]. One powerful recommendation among these argues that the potential to uncover the health effects of racism is unsacrificable. *"To truly get to the bottom of racism and its negative impact on persons of African ancestry,"* wrote Dawson in 2003, *"we cannot solve the problem by pretending that we are not what we are, or who we are, because some, be they friend or foe, are ashamed of their actions, if possible, or that of others. Because in effect, to argue that Blacks should somehow desist from categorizing ourselves as we see fit; in our intellectual exchanges is in and of itself, in the final analysis, a pernicious form of racism and self-hatred if that is the source of calls for us to 'change.'"* [104]

There are some other recommendations that justify the use of population categories in research, but only when researchers take additional measures to minimize confusion, unsupported conclusions, and arbitrary employment of population categories [23,24,62,78,107,108]. These measures include collecting data on socioeconomic status and genetics to supplement analysis, precisely defining employed population categories, and establishing a specific health or scientific interest furthered by the use of categorization, but some define these measures with sweeping statements like "tak[ing] great care." [107] Others justify the use of population categories only for research with certain purposes, such as to survey incidence of disease, to examine the interface between social and biological variation, or to lay the ground for further research into more specific risk factors. [39,52,95,106,109] The last four recommendations then urge investigators and authors to be sensitive with the terminology used to describe populations in such research, in order to avoid confusion and the insinuation of a hierarchy between groups. [21,32,44]

However, some recommendations completely discourage the employment of certain population categories in research. Overall, these pertain entirely to race and ethnicity, and none broadly call for ancestry not to be employed in research. Two recommendations within this sub-theme call for the refusal of approval and/or funding to genetic studies using race as a variable. [101,108] Other recommendations urge clinicians and researchers to minimize or eliminate their use of race and ethnicity [21,25,36,68,110], to focus their efforts on genetic variation [111], or to focus on ethnic rather than racial categorization [29]. Finally, ten more recommendations within this theme discourage researchers from using race as a proxy for other variables, most often ancestry, in their research. [27,32,39,70,112–115] Six more recommendations urge investigators to collect ancestry data instead of race or ethnicity, in order to better understand the variation captured by their study participant cohorts or to investigate a genetic basis for any observed disparities. [25,98,101,102,108] Only one article fundamentally problematizes the use of ancestry in medicine, but only in the context of developing individualized treatments: the authors state that race and ancestry are both *"imperfect measures that will both slow researchers' progress toward individualized genetic knowledge and provide clinicians with incomplete, potentially misleading information."* [116]

*8. The need for further research and guidelines*

Fifty-seven (17%) extracted recommendations fall under this theme. Most of these encourage further research into several different aspects of and important considerations regarding the use of race, ethnicity, and ancestry in health and sciences research. This includes research to elucidate where population categories lie along the social-biological axis [28,35,41,55,81,117,118] and further investigation of existing disparities between populations to identify their causes and to address them [84,92,119]. However, this subcategory is also marked by a high incidence of ancestry-related recommendations, which state the need to optimize the utility and robustness of genetic ancestry estimation [11,43,120,121] and to learn how to best characterize human genetic variation, whether that be by continental ancestry categories or not. [118,122]

The rest of the recommendations within this theme call for the development of further guidelines and regulations for future research involving race, ethnicity, and/or ancestry. Many of these are in vague terms, calling for the formation of policies or new definitions that "mitigate risks" or "address challenges". [15,19,35,74,77,81,109,117,123] However, many recommendations call for journals or centers to adopt the specific standards or guidelines suggested by other articles in this collection, such as those that suggest a need for transparency. [15,33,46,56,89,122,124] One article contains the ancestry-specific recommendations that pertain to this category; it urges leaders in the human genetics field to convene to develop standards of practice and communication of ancestry-testing and research, including standards of representing statistical confidence in ancestry estimation and guidelines for appropriate terminology. [120]

**DISCUSSION**

This systematic review analyzed a total of 121 articles published since 2000 that provide normative recommendations on the use of race, ethnicity, and ancestry in science, medicine, and public health, with particular focus on genomic research. Recommendations were consistently published throughout the 21-year period analyzed (see Figure 2a), demonstrating an ongoing perception of the need for such normative criteria. Articles appeared in journals from a wide variety of domains (see Figure 2b), demonstrating that this perceived need extends across several disciplines.

Most if not all contributing authors for 84% of these articles were affiliated with the USA. The dominance of articles from the United States may be a result of its particularly dark history of systemic racism motivating exceptional concern for rooting out notions of prejudice and hierarchy from customs of population categorization. Further, a cross-national survey of the 2000 global census, published in 2008, revealed that of all 147 national census questionnaires examined, only that of the United States measured race and ethnicity separately. [125] As the report states: *"In this view, which is extremely unusual in international perspective, ethnic groups are different from races because they are rooted in sociohistorical contexts; races thus appear to be grounded in something other than social processes."* This practice may contribute to general confusion within the United States about the significance of these categories, and by extension the context for their usage in medicine and science.

Seven out of eight themes identified during analysis were marked by significant consensus amongst the recommendations within them, indicating that there are certain uncontroversial sources of concern but that these have remained unresolved throughout the analyzed time period. Even agreeing recommendations, however, broadly give a sense of confusion and frustration. Different recommendations pertaining to the same theme often drew attention to the same issue but proposed different approaches. For example, some articles that drew attention to the need for transparency urged researchers to be transparent about different aspects of their research—while some focused on the need for clear definitions of terms like "race" and ethnicity", others encouraged transparency about research methods, like how and by whom study participants are categorized by race or ethnicity. Furthermore, the entire set of recommendations pertaining to transparency demonstrated pervasive frustration with vague circulating definitions and practices that can lead to subpar or nonexistent justifications for the employment of population categories in research.

Finally, the third most popular theme calls for further research to be conducted and further guidelines to be established about the use of race, ethnicity, and ancestry in science, medicine, and public health. The prevalence of recommendations within this theme illustrates a clear desire to pave the way for future normative criteria that are more informed and organized than the existing set of recommendations.

In contrast to the seven themes with broad agreement, the theme with the most recommendations, *Appropriate definitions of population categories and contexts for use*, contained recommendations that significantly disagreed with one another. Many recommendations defined "race" at wildly different places along the social-biological axis, and many put forth differing opinions on how strictly population categories in general should be defined, ranging from proposals strictly delineating the definitions of race and ethnicity [67], to the proposal that they be jointly be defined as *"any classification, whether genetic or self-reported, cognizable and supported by evidence in medical, legal, or social disciplines."* [52]

The recommendations specifically pertaining to the *use* of population categories then argued one of three disparate positions: that it is always acceptable to use certain population categories in research, that it is only acceptable to use them under certain conditions, and that it is never acceptable to use them. The considerable disagreement amongst these recommendations indicates that a lack of clear, centralized definitions of population categories goes hand in hand with confusion and disagreement about when they should be used. The large volume of recommendations within this theme then arguably indicates that answering questions about the definition and appropriate use of population categories will be foundational to approaching the problems identified throughout all themes.

Thirty-four of the 121 examined articles and 60 of the 334 extracted recommendations pertain to ancestry. Most of the recommendations about ancestry relate to establishing its relationship with the concepts of race and ethnicity. Many argue for an active differentiation of race and ethnicity from ancestry, especially by direct-to-consumer ancestry companies, in order to avoid misinterpretation of the meaning and significance of ancestral differences by the public. However, some disagree about the nature of the relationship between ancestry data and racial data. For example, while certain recommendations encourage researchers not to use racial data as a proxy for ancestral data or even support using ancestral data in place of racial data, others argue that there is significant overlap between racial and ancestral categories, so racial data should be collected when ancestral data is too costly to obtain. This disagreement, along with the many recommendations in *The need for further research and guidelines* theme that encourage further research into how to accurately estimate and describe genetic ancestry, illustrate that a clear role for and understanding of the value of ancestry in science and medicine is yet to be established. This imperfect understanding of ancestry, its significance, and how it should be used is also reflected in the many recommendations that call for investigators to adjust their research methods in order to avoid overestimating the impact of genetic ancestry on their results. Several of these urge researchers and care providers to consistently include an acknowledgement of other variables, such as socioeconomic status, when investigating or considering ancestry.

However, few of the extracted recommendations focus on the potential dangers of using ancestry categories. Only one fundamentally problematizes the use of ancestry, and only in the context of precision medicine.[116] Recommendations in *The use of appropriate statistical methodologies* theme that are critical of the use of ancestry propose amendments to the statistical methodologies currently used to infer ancestry, and these only problematize the way ancestry is estimated, not the contexts in which it is used. Recommendations in *The need for diverse samples and practitioners* theme only criticize the overrepresentation of European samples in genomic research in particular, and urge researchers to improve the applicability of such research by collecting samples from a broader range of individuals. Furthermore, one of the most-cited articles in this collection, "The meanings of "race" in the new genomics: implications for health disparities research" by Lee et al. in 2001, calls ancestry a *"neutral word"* that can be used to *"avoid potentially misleading terms."*[78] Overall, the extracted body of recommendations does not demonstrate the same examination of the acceptability of the use of ancestry that it does for race and ethnicity. This begs the question: is ancestry being touted as a cure-all to the problems raised by the use of race and ethnicity in science and medicine? The objective value and potential dangers of the use of ancestry thus provide potential areas of examination for future normative research that will rely heavily on the engagement of the field of genetics in questions about the value of population descriptors.

**CONCLUSION**

We have conducted the first systematic review of normative guidelines pertaining to the use of population categories in science, medicine, and public health, with particular focus on genomic research. Examined recommendations were published consistently throughout the examined time period and in a wide range of journals, illustrating a persistent and interdisciplinary recognition of the need for normative guidance on this topic.

We identified seven themes with broad agreement across recommendations: *The need for transparency; Awareness of impact on particular communities; The use of appropriate statistical methodologies; The role of public reactions and public engagement; The need for diverse samples and practitioners; The need for an appreciation of nuance; The need for further research and guidelines*. Despite broadly agreeing, however, recommendations pertaining to these themes reflected a sense of confusion, uncertainty, and persistent dissatisfaction. Many used nonspecific language to identify pertinent issues without proposing concrete solutions; others proposed different solutions to the same issues. Recommendations under the theme of *The need for transparency* conveyed particular frustration at the current lack of explicit definitions for many of the population categories in circulation. We identified significant fundamental disagreement within one theme: *Appropriate definitions of population categories and contexts for use.* Finally, we identified that although many recommendations focus on the inappropriate use of race and ethnicity, and some even condemn their usage entirely, none fundamentally problematize the use of ancestry, and criticism of ancestry is limited to calls to refine the statistical strength of ancestral estimations and to improve the applicability of genetic research.

This review demonstrates a strong need for new, clear, and effective guidelines on the use of population categories in science, medicine, and public health, and a particular need to investigate the appropriate definition and use of ancestry within these disciplines. The applicability of these guidelines should extend across disciplinary and national borders, given the global and interdisciplinary concern about this topic represented in this examined pool of recommendations, and should particularly engage the genetics field as they develop new standards for describing between-group differences.[8]

**SUPPLEMENTARY INFORMATION**

One excel file containing information on all the articles meeting inclusion criteria, all extracted recommendations, and theme assignments to these recommendations. (Please email annalewis@fas.harvard.edu for a copy).

**ACKNOWLEDGEMENTS**

We thank the members of the E J Safra working group on the concepts of ancestry, genetic ancestry, and population for their input on this work.

**DECLARATIONS OF INTEREST**

B.M.N. is a member of the scientific advisory board at Deep Genomics and RBNC Therapeutics, a member of the scientific advisory committee at Milken, and a consultant for Camp4 Therapeutics and Merck. A.C.F.L. owns stock in Fabric Genomics.

**FIGURES**

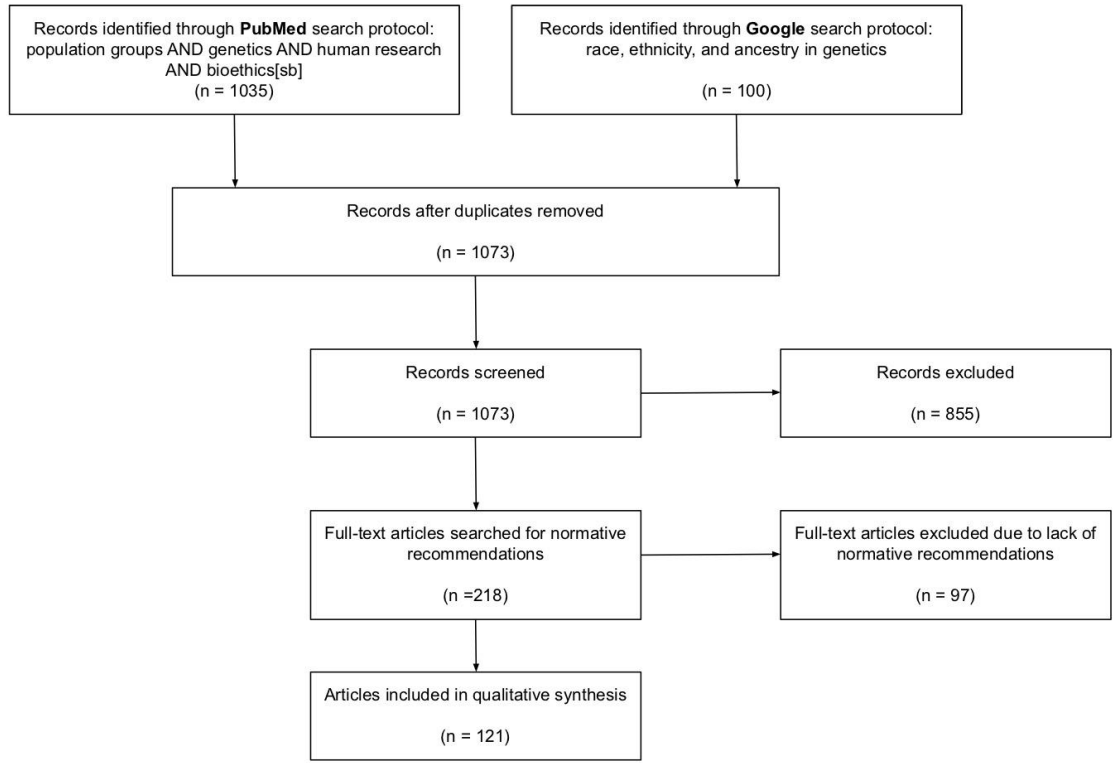

**Figure 1: PRISMA Diagram, demonstrating the process of article inclusion and exclusion.** 121 articles returned by the complete search strategy yielded 334 normative recommendations on the use of race, ethnicity, and ancestry in science, medicine, and public health.

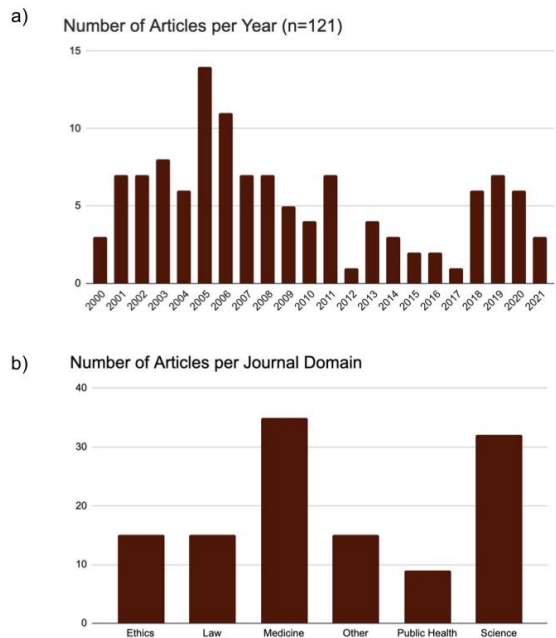

**Figure 2: Articles continuing normative recommendations on the use of population categories have been published consistently since the year 2000, and across a broad variety of domains.** a) Although there is a clear clustering of articles around the first decade of the 21st century, articles pertaining to this topic have been published consistently throughout the 21 examined years. b) While most articles hail from journals with a majorly scientific or medical focus, the other listed domains are also well-represented, suggesting that this topic is truly an interdisciplinary one. Journals categorized within the "Other" domain have main focuses ranging from history to philosophy (see Supplementary Materials).

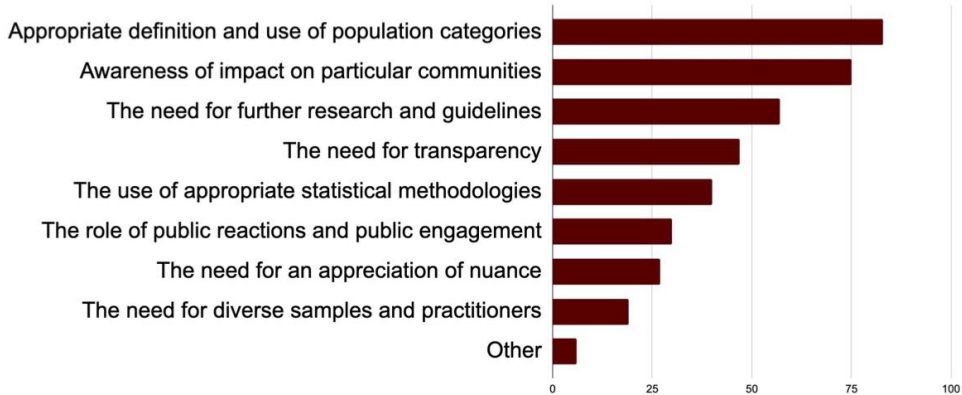

**Figure 3: Extracted recommendations were distributed across eight themes.** Before these eight themes were identified, each recommendation was examined and tagged based on its content—these twenty-nine tagged "sub-themes" were then sorted into eight larger themes. Those which did not correspond to a sub-theme were labeled "Other". Table 1 reports the sub-themes that correspond to each theme.

**TABLES**

**Table 1: Common themes and sub-themes identified among included articles**

| Theme | Sub-themes |
|---|---|
| The need for transparency | a) Investigators should be transparent about why they are using population categories. b) Investigators should be transparent about how any population categories used are defined. c) Investigators should make their data and collection methods publicly accessible. d) Investigators should fully analyze and explain their findings. |
| Awareness of | a) Investigators should respect study populations. |

| impact on particular communities | b) Awareness of past and present impacts of racism in science and medicine should be encouraged/increased. c) Investigators should consult specific communities during the research process. |
|---|---|
| The use of appropriate statistical methodologies | a) Investigators should not make causal claims based on associations found in their data. b) Relates to the statistical interpretation of genetic data. c) Investigators should be aware of the contribution of factors other than race/ethnicity, such as socioeconomic circumstances or ancestry, to health outcomes. d) Investigators should match collected variables to their study question(s). |
| The role of public reactions and public engagement | a) Investigators should be aware of the potential social impact of research involving population categories, and sensitively release results of or information about this research to the public. b) Investigators and institutions should consult and educate the public. c) The scientific community should increase visibility of and education about important considerations for research involving race, ethnicity, and/or ancestry. |
| The need for diverse samples and practitioners | a) Investigators should recruit more diverse cohorts of study participants. b) The makeup of and discussions within the medical community should be diversified. c) Researchers should ensure that study populations are representative of the investigated population. |
| The need for an appreciation of nuance | a) Those in the scientific or medical field need to use and consider race/ethnicity in a more nuanced fashion. b) Those in the scientific or medical field need to use and consider ancestry/population in a more nuanced fashion. c) Those in the scientific or medical field need to consider the relationship between race, ethnicity, ancestry, and population in a more nuanced fashion. |
| Appropriate definitions of population categories and contexts for use | a) Certain population categories should not be employed in research.<br>b) The scientific/medical community should continue collecting race and/or ethnicity data and investigating race- or ethnicity-based disparities.<br>c) It is appropriate for investigators to employ population categories under certain conditions.<br>d) Researchers should adopt or avoid certain definitions of population categories.<br>e) Investigators should collect ancestry data instead of race or ethnicity.<br>f) Race should not be used as a proxy for other variables in research.<br>g) Investigators and authors should be sensitive with terminology. |
| The need for further research and guidelines | a) Guidelines should be developed and regulation tightened for future research involving race, ethnicity, or ancestry.<br>b) Further research into important considerations and current pitfalls of research involving race, ethnicity, and/or ancestry should be conducted. |

## REFERENCES

1. Lopez, L., III, Hart, L.H., III, and Katz, M.H. (2021). Racial and Ethnic Health Disparities Related to COVID-19. JAMA *325*, 719–720.
2. Zavala, V.A., Bracci, P.M., Carethers, J.M., Carvajal-Carmona, L., Coggins, N.B., Cruz-Correa, M.R., Davis, M., de Smith, A.J., Dutil, J., Figueiredo, J.C., et al. (2021). Cancer health disparities in racial/ethnic minorities in the United States. Br. J. Cancer *124*, 315–332.
3. Burton, D.C., Flannery, B., Bennett, N.M., Farley, M.M., Gershman, K., Harrison, L.H., Lynfield, R., Petit, S., Reingold, A.L., Schaffner, W., et al. (2010). Socioeconomic and Racial/Ethnic Disparities in the Incidence of Bacteremic Pneumonia Among US Adults. Am. J. Public Health *100*, 1904–1911.
4. Forno, E., and Celedón, J.C. (2012). Health Disparities in Asthma. Am. J. Respir. Crit. Care Med. *185*, 1033–1035.
5. Graham, G. (2015). Disparities in Cardiovascular Disease Risk in the United States. Curr. Cardiol. Rev. *11*, 238–245.
6. Khazanchi, R., Evans, C.T., and Marcelin, J.R. (2020). Racism, Not Race, Drives Inequity Across the COVID-19 Continuum. JAMA Netw. Open *3*, e2019933.
7. Raghav, K., Anand, S., Gothwal, A., Singh, P., Dasari, A., Overman, M.J., and Loree, J.M.

(2021). Underreporting of race/ethnicity in COVID-19 research. Int. J. Infect. Dis. *108*, 419–421.
8. Use of Race Ethnicity and Ancestry as Population Descriptors in Genomics Research | National Academies.
9. Byeon, Y.J.J., Islamaj, R., Yeganova, L., Wilbur, W.J., Lu, Z., Brody, L.C., and Bonham, V.L. (2021). Evolving use of ancestry, ethnicity, and race in genetics research—A survey spanning seven decades. Am. J. Hum. Genet. *108*, 2215–2223.
10. Vyas, D.A., Eisenstein, L.G., and Jones, D.S. (2020). Hidden in Plain Sight — Reconsidering the Use of Race Correction in Clinical Algorithms. N. Engl. J. Med. *383*, 874–882.
11. Borrell, L.N., Elhaway, J.R., Fuentes-Afflick, E., Witonsky, J., Bhakta, N., Wu, A.H.B., Bibbins-Domingo, K., Rodríguez-Santana, J.R., Lenoir, M.A., Gavin, J.R., et al. (2021). Race and Genetic Ancestry in Medicine — A Time for Reckoning with Racism. N. Engl. J. Med. *384*, 474–480.
12. Oni-Orisan, A., Mavura, Y., Banda, Y., Thornton, T.A., and Sebro, R. (2021). Embracing Genetic Diversity to Improve Black Health. N. Engl. J. Med. *384*, 1163–1167.
13. Getting genetic ancestry right for science and society.
14. Bliss, C. (2020). Conceptualizing Race in the Genomic Age. Hastings Cent. Rep. *50 Suppl 1*, S15–S22.
15. Ali-Khan, S.E., Krakowski, T., Tahir, R., and Daar, A.S. (2011). The use of race, ethnicity and ancestry in human genetic research. HUGO J. *5*, 47–63.
16. Sankar, P., Cho, M.K., and Mountain, J. (2007). Race and Ethnicity in Genetic Research. Am. J. Med. Genet. A. *143*, 961–970.
17. Smart, A., Tutton, R., Martin, P., Ellison, G.T.H., and Ashcroft, R. (2008). The standardization of race and ethnicity in biomedical science editorials and UK biobanks. Soc. Stud. Sci. *38*, 407–423.
18. Page, M.J., McKenzie, J.E., Bossuyt, P.M., Boutron, I., Hoffmann, T.C., Mulrow, C.D., Shamseer, L., Tetzlaff, J.M., Akl, E.A., Brennan, S.E., et al. (2021). The PRISMA 2020 statement: an updated guideline for reporting systematic reviews. BMJ n71.
19. Callier, S.L. (2019). The Use of Racial Categories in Precision Medicine Research. Ethn. Dis. *29*, 651–658.
20. (2000). Census, race and science. Nat. Genet. *24*, 97–98.
21. Duggan, C.P., Kurpad, A., Stanford, F.C., Sunguya, B., and Wells, J.C. (2020). Race, ethnicity, and racism in the nutrition literature: an update for 2020. Am. J. Clin. Nutr. *112*, 1409–1414.
22. Foster, M.W. (2009). Looking for race in all the wrong places: analyzing the lack of productivity in the ongoing debate about race and genetics. Hum. Genet. *126*, 355–362.
23. Jones, C.P. (2001). Invited Commentary: "Race," Racism, and the Practice of Epidemiology. Am. J. Epidemiol. *154*, 299–304.
24. Kahn, J. (2006). Genes, Race, and Population: Avoiding a Collision of Categories. Am. J. Public Health *96*, 1965–1970.
25. Lee, S.S.-J., Mountain, J., Koenig, B., Altman, R., Brown, M., Camarillo, A., Cavalli-Sforza, L., Cho, M., Eberhardt, J., Feldman, M., et al. (2008). The ethics of characterizing difference: guiding principles on using racial categories in human genetics. Genome Biol. *9*, 404.
26. Winker, M.A. (2004). Measuring race and ethnicity: why and how? JAMA *292*, 1612–1614.
27. Winker, M.A. (2006). Race and Ethnicity in Medical Research: Requirements Meet Reality. J. Law. Med. Ethics *34*, 520–525.
28. Bhopal, R. (2006). Race and Ethnicity: Responsible Use from Epidemiological and Public Health Perspectives. J. Law. Med. Ethics *34*, 500–507.
29. Rothstein, M.A., and Epps, P.G. (2001). Pharmacogenomics and the (ir)relevance of race. Pharmacogenomics J. *1*, 104–108.
30. Sankar, P., and Cho, M.K. (2002). Toward a New Vocabulary of Human Genetic Variation.

Science *298*, 1337–1338.
31. Terry, S.F., Christensen, K.D., Metosky, S., Rudofsky, G., Deignan, K.P., Martinez, H., Johnson-Moore, P., and Citrin, T. (2012). Community Engagement about Genetic Variation Research. Popul. Health Manag. *15*, 78–89.
32. Khan, A.T., Gogarten, S.M., McHugh, C.P., Stilp, A.M., Sofer, T., Bowers, M., Wong, Q., Cupples, L.A., Hidalgo, B., Johnson, A.D., et al. (2021). Recommendations on the use and reporting of race, ethnicity, and ancestry in genetic research: experiences from the NHLBI Trans-Omics for Precision Medicine (TOPMed) program. ArXiv210807858 Q-Bio.
33. Reverby, S.M. (2010). Invoking "Tuskegee": problems in health disparities, genetic assumptions, and history. J. Health Care Poor Underserved *21*, 26–34.
34. Lee, S.S. (2009). Pharmacogenomics and the challenge of health disparities. Public Health Genomics *12*, 170–179.
35. Lee, S.S.-J. (2005). Racializing Drug Design: Implications of Pharmacogenomics for Health Disparities. Am. J. Public Health *95*, 2133–2138.
36. Takezawa, Y., Kato, K., Oota, H., Caulfield, T., Fujimoto, A., Honda, S., Kamatani, N., Kawamura, S., Kawashima, K., Kimura, R., et al. (2014). Human genetic research, race, ethnicity and the labeling of populations: recommendations based on an interdisciplinary workshop in Japan. BMC Med. Ethics *15*, 33.
37. Capocasa, M., and Volpi, L. (2019). The ethics of investigating cultural and genetic diversity of minority groups. Homo Int. Z. Vgl. Forsch. Am Menschen *70*, 233–244.
38. Claw, K.G., Anderson, M.Z., Begay, R.L., Tsosie, K.S., Fox, K., and Garrison, N.A. (2018). A framework for enhancing ethical genomic research with Indigenous communities. Nat. Commun. *9*, 2957.
39. Ossorio, P., and Duster, T. (2005). Race and genetics: Controversies in biomedical, behavioral, and forensic sciences. Am. Psychol. *60*, 115–128.
40. Clayton, E.W. (2002). Complex Relationship of Genetics, Groups, and Health: What It Means for Public Health Symposium Article - Part III: Salient Issues in Public Health Law. J. Law. Med. Ethics *30*, 290–297.
41. Brewer, R.M. (2006). Thinking Critically about Race and Genetics. J. Law. Med. Ethics *34*, 513–519.
42. Francis, C.K. (2001). The medical ethos and social responsibility in clinical medicine. J. Natl. Med. Assoc. *93*, 157–169.
43. Kittles, R.A., and Weiss, K.M. (2003). Race, ancestry, and genes: implications for defining disease risk. Annu. Rev. Genomics Hum. Genet. *4*, 33–67.
44. Rambachan, A. (2018). Overcoming the Racial Hierarchy: the History and Medical Consequences of "Caucasian." J. Racial Ethn. Health Disparities *5*, 907–912.
45. Rusert, B.M., and Royal, C.D.M. (2011). Grassroots Marketing in a Global Era: More Lessons from BiDil. J. Law. Med. Ethics *39*, 79–90.
46. Schwartz, R.S. (2001). Racial profiling in medical research. N. Engl. J. Med. *344*, 1392–1393.
47. Bhopal, R. (2009). Medicine and public health in a multiethnic world. J. Public Health Oxf. Engl. *31*, 315–321.
48. Bonham, V.L., and Knerr, S. (2008). Social and ethical implications of genomics, race, ethnicity, and health inequities. Semin. Oncol. Nurs. *24*, 254–261.
49. Tashiro, C.J. (2005). The meaning of race in health care and research--part 1: the impact of history. Pediatr. Nurs. *31*, 208–210.
50. Tashiro, C.J. (2005). The meaning of race in healthcare and research--part 2. Current controversies and emerging research. Pediatr. Nurs. *31*, 305–308.
51. Hoffman, S. (2005). "Racially-Tailored" Medicine Unraveled (Rochester, NY: Social Science Research Network).
52. Ruel, M.D. (2006). Using race in clinical research to develop tailored medications. Is the

FDA encouraging discrimination or eliminating traditional disparities in health care for African Americans? J. Leg. Med. *27*, 225–241.
53. Sharp, R.R., and Foster, M.W. (2007). Grappling with groups: protecting collective interests in biomedical research. J. Med. Philos. *32*, 321–337.
54. Hausman, D. (2008). Protecting groups from genetic research. Bioethics *22*, 157–165.
55. Ilkilic, I., and Paul, N.W. (2009). Ethical aspects of genome diversity research: genome research into cultural diversity or cultural diversity in genome research? Med. Health Care Philos. *12*, 25–34.
56. Sharp, R.R., and Foster, M.W. (2002). An analysis of research guidelines on the collection and use of human biological materials from American Indian and Alaskan Native communities. Jurimetrics *42*, 165–186.
57. Bankoff, R.J., and Perry, G.H. (2016). Hunter-gatherer genomics: Evolutionary insights and ethical considerations. Curr. Opin. Genet. Dev. *41*, 1–7.
58. Boyer, B.B., Dillard, D., Woodahl, E.L., Whitener, R., Thummel, K.E., and Burke, W. (2011). Ethical issues in developing pharmacogenetic research partnerships with American Indigenous communities. Clin. Pharmacol. Ther. *89*, 343–345.
59. Hiratsuka, V.Y., Hahn, M.J., Woodbury, R.B., Hull, S.C., Wilson, D.R., Bonham, V.L., Dillard, D.A., Avey, J.P., Beckel-Mitchener, A.C., Blome, J., et al. (2020). Alaska Native genomic research: perspectives from Alaska Native leaders, federal staff, and biomedical researchers. Genet. Med. *22*, 1935–1943.
60. McGregor, J.L. (2007). Population genomics and research ethics with socially identifable groups. J. Law Med. Ethics J. Am. Soc. Law Med. Ethics *35*, 356–370.
61. Foster, M.W., and Sharp, R.R. (2000). Genetic research and culturally specific risks: one size does not fit all. Trends Genet. TIG *16*, 93–95.
62. Foster, M.W., and Sharp, R.R. (2002). Race, ethnicity, and genomics: social classifications as proxies of biological heterogeneity. Genome Res. *12*, 844–850.
63. Fullerton, S.M., Yu, J.-H., Crouch, J., Fryer-Edwards, K., and Burke, W. (2010). Population description and its role in the interpretation of genetic association. Hum. Genet. *127*, 563–572.
64. Race, Ethnicity, and Genetics Working Group (2005). The use of racial, ethnic, and ancestral categories in human genetics research. Am. J. Hum. Genet. *77*, 519–532.
65. Barr, D.A. (2005). The Practitioner's Dilemma: Can We Use a Patient's Race To Predict Genetics, Ancestry, and the Expected Outcomes of Treatment? Ann. Intern. Med. *143*, 809–815.
66. Matthews-Juarez, P., and Juarez, P.D. (2011). Cultural competency, human genomics, and the elimination of health disparities. Soc. Work Public Health *26*, 349–365.
67. Mersha, T.B., and Abebe, T. (2015). Self-reported race/ethnicity in the age of genomic research: its potential impact on understanding health disparities. Hum. Genomics *9*, 1.
68. Bonham, V.L., Green, E.D., and Pérez-Stable, E.J. (2018). Examining How Race, Ethnicity, and Ancestry Data Are Used in Biomedical Research. JAMA *320*, 1533–1534.
69. Whittle, P.M. (2010). Health, inequality and the politics of genes. N. Z. Med. J. *123*, 67–75.
70. Batai, K., Hooker, S., and Kittles, R.A. (2021). Leveraging genetic ancestry to study health disparities. Am. J. Phys. Anthropol. *175*, 363–375.
71. Braun, L. (2002). Race, Ethnicity, and Health: Can Genetics Explain Disparities? Perspect. Biol. Med. *45*, 159–174.
72. Egalité, N., Ozdemir, V., and Godard, B. (2007). Pharmacogenomics research involving racial classification: qualitative research findings on researchers' views, perceptions and attitudes towards socioethical responsibilities. Pharmacogenomics *8*, 1115–1126.
73. Lewontin, R. (2005). The fallacy of racial medicine: confusions about human races. Genewatch Bull. Comm. Responsible Genet.
74. Martschenko, D.O., and Smith, M. (2021). Genes do not operate in a vacuum, and neither should our research. Nat. Genet. *53*, 255–256.

75. Palsson, G., and Helgason, A. (2003). Blondes, lost and found: representations of genes, identity, and history. Dev. World Bioeth. *3*, 159–169.

76. Balaban, E. (2005). The new racial economy: making a silk purse out of the sow's ear of racial distinctions. Genewatch Bull. Comm. Responsible Genet. *18*, 8–10.

77. Lee, S.S.-J. (2003). Race, Distributive Justice and the Promise of Pharmacogenomics. Am. J. Pharmacogenomics *3*, 385–392.

78. Lee, S.S., Mountain, J., and Koenig, B.A. (2001). The meanings of "race" in the new genomics: implications for health disparities research. Yale J. Health Policy Law Ethics *1*, 33–75.

79. Parrott, R.L., Silk, K.J., Dillow, M.R., Krieger, J.L., Harris, T.M., and Condit, C.M. (2005). Development and validation of tools to assess genetic discrimination and genetically based racism. J. Natl. Med. Assoc. *97*, 980–990.

80. Walajahi, H., Wilson, D.R., and Hull, S.C. (2019). Constructing identities: the implications of DTC ancestry testing for tribal communities. Genet. Med. Off. J. Am. Coll. Med. Genet. *21*, 1744–1750.

81. Ozdemir, V., Graham, J.E., and Godard, B. (2008). Race as a variable in pharmacogenomics science: from empirical ethics to publication standards. Pharmacogenet. Genomics *18*, 837–841.

82. Bloche, M.G. (2004). Race-Based Therapeutics. N. Engl. J. Med. *351*, 2035–2037.

83. Popejoy, A.B., Crooks, K.R., Fullerton, S.M., Hindorff, L.A., Hooker, G.W., Koenig, B.A., Pino, N., Ramos, E.M., Ritter, D.I., Wand, H., et al. (2020). Clinical Genetics Lacks Standard Definitions and Protocols for the Collection and Use of Diversity Measures. Am. J. Hum. Genet. *107*, 72–82.

84. Smart, A., Martin, P., and Parker, M. (2004). Tailored medicine: whom will it fit? The ethics of patient and disease stratification. Bioethics *18*, 322–342.

85. Aldhous, P. (2002). Geneticist fears "race-neutral" studies will fail ethnic groups. Nature *418*, 355–356.

86. Burchard, E.G. (2014). Medical research: Missing patients. Nature *513*, 301–302.

87. Stevens, J. (2003). Racial Meanings and Scientific Methods: Changing Policies for NIH-Sponsored Publications Reporting Human Variation. J. Health Polit. Policy Law *28*, 1033–1088.

88. Nugent, A., Conatser, K.R., Turner, L.L., Nugent, J.T., Sarino, E.M.B., and Ricks-Santi, L.J. (2019). Reporting of race in genome and exome sequencing studies of cancer: a scoping review of the literature. Genet. Med. Off. J. Am. Coll. Med. Genet. *21*, 2676–2680.

89. Popejoy, A.B., Ritter, D.I., Crooks, K., Currey, E., Fullerton, S.M., Hindorff, L.A., Koenig, B., Ramos, E.M., Sorokin, E.P., Wand, H., et al. (2018). The clinical imperative for inclusivity: Race, ethnicity, and ancestry (REA) in genomics. Hum. Mutat. *39*, 1713–1720.

90. Peterson-Iyer, K. (2008). Pharmacogenomics, ethics, and public policy. Kennedy Inst. Ethics J. *18*, 35–56.

91. Angel, R.J. (2011). Agency versus structure: genetics, group membership, and a new twist on an old debate. Soc. Sci. Med. 1982 *73*, 632–635.

92. Winkelmann, B.R. (2003). Pharmacogenomics, genetic testing and ethnic variability: tackling the ethical questions. Pharmacogenomics *4*, 531–535.

93. Garte, S. (2002). The racial genetics paradox in biomedical research and public health. Public Health Rep. *117*, 421–425.

94. Anomaly, J. (2017). Race Research and the Ethics of Belief. J. Bioethical Inq. *14*, 287–297.

95. Foster, M.W., and Sharp, R.R. (2004). Beyond race: towards a whole-genome perspective on human populations and genetic variation. Nat. Rev. Genet. *5*, 790–796.

96. Feller, L., Ballyram, R., Meyerov, R., Lemmer, J., and Ayo-Yusuf, O.A. (2014). Race/ethnicity in biomedical research and clinical practice. SADJ J. South Afr. Dent. Assoc. Tydskr. Van Suid-Afr. Tandheelkd. Ver. *69*, 272–274.

97. Frank, R. (2008). Functional or futile?: the (in)utility of methodological critiques of genetic research on racial disparities in health. A commentary on Kaufman's "Epidemiologic analysis of racial/ethnic disparities: some fundamental issues and a cautionary example." Soc. Sci. Med. 1982 *66*, 1670–1674.

98. Sade, R.M. (2007). What's right (and wrong) with racially stratified research and therapies. J. Natl. Med. Assoc. *99*, 693–696.

99. Fang, H., Hui, Q., Lynch, J., Honerlaw, J., Assimes, T.L., Huang, J., Vujkovic, M., Damrauer, S.M., Pyarajan, S., Gaziano, J.M., et al. (2019). Harmonizing Genetic Ancestry and Self-identified Race/Ethnicity in Genome-wide Association Studies. Am. J. Hum. Genet. *105*, 763–772.

100. Hardimon, M.O. (2013). Race concepts in medicine. J. Med. Philos. *38*, 6–31.

101. Payne, P.W., and Royal, C. (2007). The Role of Genetic and Sociopolitical Definitions of Race in Clinical Trials. J. Am. Acad. Orthop. Surg.

102. Shields, A.E., Fortun, M., Hammonds, E.M., King, P.A., Lerman, C., Rapp, R., and Sullivan, P.F. (2005). The use of race variables in genetic studies of complex traits and the goal of reducing health disparities: A transdisciplinary perspective. Am. Psychol. *60*, 77–103.

103. Burchard, E.G., Ziv, E., Coyle, N., Gomez, S.L., Tang, H., Karter, A.J., Mountain, J.L., Pérez-Stable, E.J., Sheppard, D., and Risch, N. (2003). The importance of race and ethnic background in biomedical research and clinical practice. N. Engl. J. Med. *348*, 1170–1175.

104. Dawson, G. (2003). Human genome, race and medicine. J. Natl. Med. Assoc. *95*, 309–312.

105. Shanawani, H., Dame, L., Schwartz, D.A., and Cook-Deegan, R. (2006). Non-reporting and inconsistent reporting of race and ethnicity in articles that claim associations among genotype, outcome, and race or ethnicity. J. Med. Ethics *32*, 724–728.

106. Kaufman, J.S., and Cooper, R.S. (2001). Commentary: Considerations for Use of Racial/Ethnic Classification in Etiologic Research. Am. J. Epidemiol. *154*, 291–298.

107. Kahn, J. (2005). Ethnic drugs. Hastings Cent. Rep. *35*, 1 p following 48.

108. Lillquist, E., and Sullivan, C.A. (2006). Legal regulation of the use of race in medical research. J. Law Med. Ethics J. Am. Soc. Law Med. Ethics *34*, 535–551, 480.

109. Jaja, C., Gibson, R., and Quarles, S. (2013). Advancing Genomic Research and Reducing Health Disparities: What Can Nurse Scholars Do? J. Nurs. Scholarsh. *45*, 202–209.

110. Umek, W., and Fischer, B. (2020). We Should Abandon "Race" as a Biological Category in Biomedical Research. Female Pelvic Med. Reconstr. Surg. *26*, 719–720.

111. Lorusso, L. (2011). The justification of race in biological explanation. J. Med. Ethics *37*, 535–539.

112. Eichelberger, K.Y., Alson, J.G., and Doll, K.M. (2018). Should Race Be Used as a Variable in Research on Preterm Birth? AMA J. Ethics *20*, 296–302.

113. Hunt, L.M., Truesdell, N.D., and Kreiner, M.J. (2013). Genes, race, and culture in clinical care: racial profiling in the management of chronic illness. Med. Anthropol. Q. *27*, 253–271.

114. Jones, T., and Roberts, J.L. (2020). GENETIC RACE? DNA ANCESTRY TESTS, RACIAL IDENTITY, AND THE LAW. Columbia Law Rev. *120*, 1929–2016.

115. Schaefer, G.O., Tai, E.S., and Sun, S.H.-L. (2020). Navigating conflicts of justice in the use of race and ethnicity in precision medicine. Bioethics *34*, 849–856.

116. Jones, D.S., and Perlis, R.H. (2006). Pharmacogenetics, race, and psychiatry: prospects and challenges. Harv. Rev. Psychiatry *14*, 92–108.

117. Craddock Lee, S.J. (2005). The Risks of Race in Addressing Health Disparities. Hastings Cent. Rep. *35*, 1p–48.

118. Lee, S.S.-J. (2007). The ethical implications of stratifying by race in pharmacogenomics. Clin. Pharmacol. Ther. *81*, 122–125.

119. Cohn, J.N. (2006). The Use of Race and Ethnicity in Medicine: Lessons from the African-American Heart Failure Trial. J. Law. Med. Ethics *34*, 552–554.
32

120. Royal, C.D., Novembre, J., Fullerton, S.M., Goldstein, D.B., Long, J.C., Bamshad, M.J., and Clark, A.G. (2010). Inferring genetic ancestry: opportunities, challenges, and implications. Am. J. Hum. Genet. *86*, 661–673.

121. Slabbert, N., and Heathfield, L.J. (2018). Ethical, legal and social implications of forensic molecular phenotyping in South Africa. Dev. World Bioeth. *18*, 171–181.

122. Zhang, F., and Finkelstein, J. (2019). Inconsistency in race and ethnic classification in pharmacogenetics studies and its potential clinical implications. Pharmacogenomics Pers. Med. *12*, 107–123.

123. Lee, S.S.-J., Soo-Jin Lee, S., Bolnick, D.A., Duster, T., Ossorio, P., and Tallbear, K. (2009). Genetics. The illusive gold standard in genetic ancestry testing. Science *325*, 38–39.

124. Bamshad, M. (2007). Lost in translation: Meaningful policies for writing about genetics and race. Am. J. Med. Genet. A.

125. Morning, A. (2015). Ethnic Classification in Global Perspective: A Cross-National Survey of the 2000 Census Round. In Social Statistics and Ethnic Diversity: Cross-National Perspectives in Classifications and Identity Politics, P. Simon, V. Piché, and A.A. Gagnon, eds. (Cham: Springer International Publishing), pp. 17–37.